\documentclass[aps,prl,twocolumn,superscriptaddress,floatfix]{revtex4-2}
\usepackage{amsmath,amsfonts,amssymb,bm}
\usepackage{graphicx}
\usepackage[usenames]{color}
\usepackage{hyperref,epsfig,wrapfig}

\begin{document}

\title{Fully-frustrated octahedral antiferromagnets: emergent complexity in external field}

\author{A. S. Gubina}
\affiliation{Universit\'e Grenoble Alpes, CEA, IRIG, PHELIQS, 38000 Grenoble, France}
\author{T. Ziman}
\affiliation{Institut Laue Langevin, 38042 Grenoble Cedex 9, France}
\affiliation{Kavli Institute for Theoretical Science, University of the Chinese Academy of Sciences,
100190, Beijing,  China}
 
\author{M. E. Zhitomirsky}
\affiliation{Universit\'e Grenoble Alpes, CEA, IRIG, PHELIQS, 38000 Grenoble, France}
\affiliation{Institut Laue Langevin, 38042 Grenoble Cedex 9, France}

\date{\today}
             
\begin{abstract}
Octahedral antiferromagnets are distinguished by crystal lattices composed of octahedra of magnetic ions.
In the fully frustrated case, the Heisenberg Hamiltonian can be represented as a sum of squares of total spins 
for each octahedral block. We study the fully frustrated spin model for a lattice of edge-shared octahedra, 
which corresponds to  the $J_1$--$J_2$ fcc antiferromagnet with $J_2/J_1 = 1/2$. The magnetization 
process at this strongly frustrated point features a remarkably rich sequence of different magnetic phases 
that include fractional plateaus at $m = 1/3$ and $2/3$ values of the total magnetization. By performing extensive 
Monte Carlo simulations we construct the $H$--$T$ phase diagram of the classical model with eight field-induced 
states, which acquire stability via the order by disorder mechanism. These antiferromagnetic states have distinct spin configurations of their octahedral blocks.  The same spin configurations are also relevant for the fully frustrated corner-shared model bringing an apparent similarity to their field-induced states.
\end{abstract}

\maketitle

\textit{Introduction.}\ ---
Geometrical frustration in a magnetic material occurs when the lattice topology prevents simultaneous minimization
of all pairwise interactions between local moments. Frustrated spin models may possess accidental degeneracies,
which lead to enhanced low-temperature fluctuations and, as a result, to a variety of new complex states \cite{WS94,Springer11,Zhou17,Savary17,Broholm20}. Experimental and theoretical studies  have 
mostly focused  on lattices based on triangular or tetrahedral spin units such as triangular, kagome, and
pyrochlore antiferromagnets and their descendants \cite{Springer11}.

\begin{figure}[tb]
\centerline{\includegraphics[width=0.9\columnwidth]{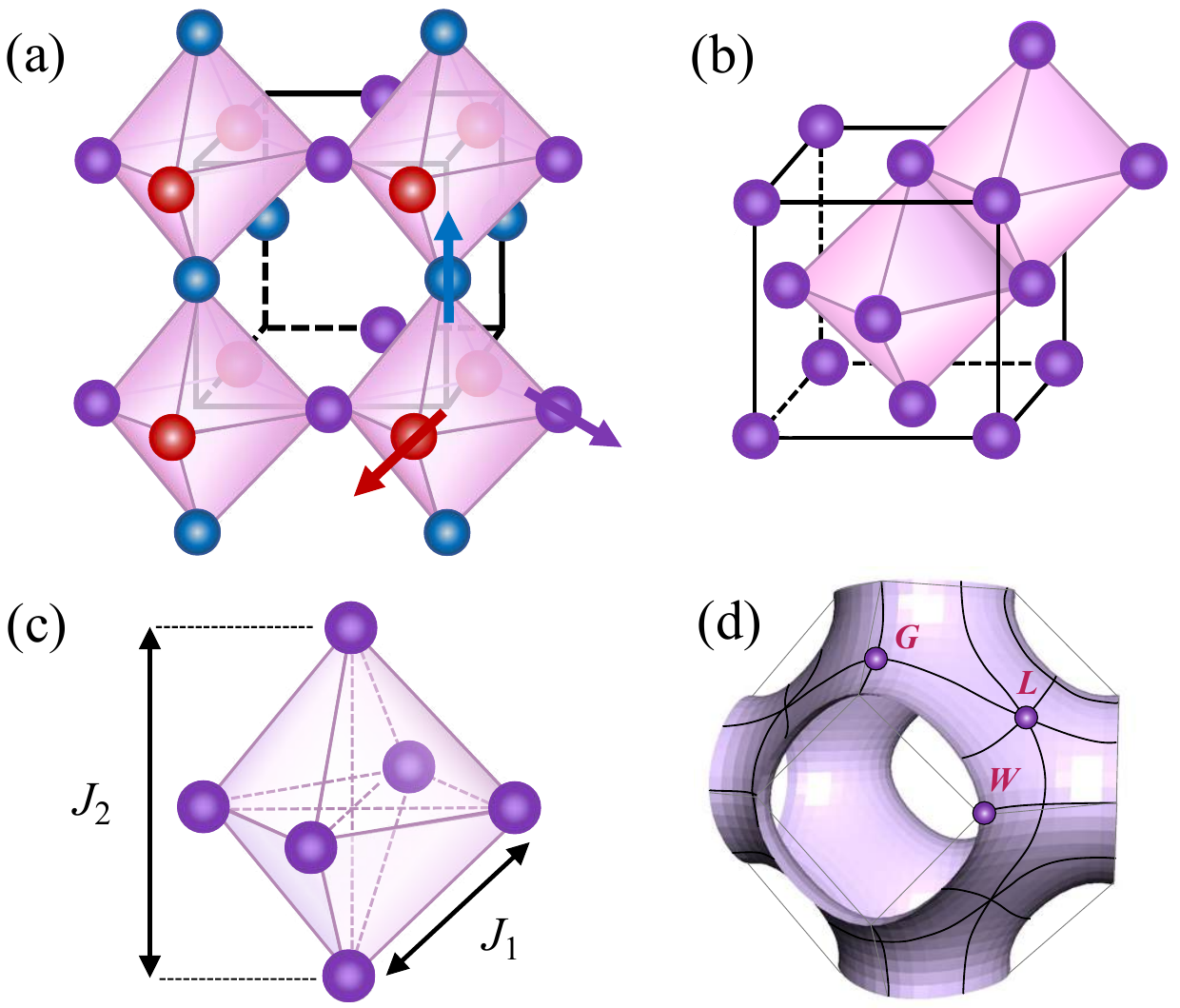}}
\caption{Examples of octahedral lattices. (a) Corner-shared network realized in cubic antiperovskites. 
Colored spheres specify the $q=0$  antiferromagnetic state with the $120^\circ$  
arrangement of spins.
(b) Edge-shared network emerging in the $J_1$--$J_2$ fcc antiferromagnet. (c) A spin octahedron with 
two types of exchange interactions. (d) Manifold of degenerate wavevectors corresponding to the minimum of $J(\bm{q})$ 
in the Brillouin zone for the fcc model with $J_2/J_1 = 1/2$.  Special symmetry points: $L$ or 
$\bm{Q}_2 = (\pi,\pi,\pi)$, 
$W$ or  $\bm{Q}_3 = ( 2\pi, \pi, 0)$,  $G$ or  $\bm{Q}_4=(4\pi/3,  0, 4\pi/3)$.}
    \label{fig:geometry}
\end{figure}

Octahedral antiferromagnets constitute a distinct class of geometrically frustrated magnets that has  
received only limited attention to date. Magnetic ions in these materials form lattices composed of octahedral
blocks. Two of such structures formed by corner- and edge-shared octahedra are 
presented in Figs.~\ref{fig:geometry}(a) and \ref{fig:geometry}(b), respectively. The first type of magnetic 
lattice is realized in antiperovskites \cite{Deng23,Wang20,Singh18} and Mn$_3$X ordered alloys \cite{Kren68,Tomeno99}. The edge-shared octahedra are inherent to the face-centered cubic (fcc) structure 
and appear as elementary frustrated units in the $J_1$--$J_2$ antiferromagnetic fcc materials \cite{Seehra88}
at a fully frustrated point $J_2/J_1 = 1/2$, as detailed below. Recently, there have been a number of 
experimental \cite{Takenaka09,Matsuoka18,Hirschmann22,Li23,Franco24}
and theoretical \cite{Tahara07,Hemmati12,Sklan13,LeBlanc13,Benton21,Szabo22,Paddison24} studies 
of octahedral materials and models with a focus on their unusual magnetic properties. In particular, 
complex $H$--$T$ phase diagrams have been measured for europium antiperovskites 
Eu$_3$PbO \cite{Hirschmann22}  and  
Eu$_3$SnO \cite{Li23}. At low temperatures,  their magnetization curves exhibit several plateau-like features, 
including one close to 1/3 of the total magnetization.  Magnetic Eu$^{2+}$ ions in both compounds have 
large semiclassical spins $S=7/2$, typically with a weak magnetic anisotropy. Hence, the magnetization 
plateaus in these materials can result from magnetic frustration in the octahedral lattice.

The frustrated units or blocks that make up a lattice influence the magnetization process 
of the corresponding antiferromagnetic model. In particular, the semi-classical magnetization plateaus 
depend on the number of spins in each block \cite{Zhitomirsky00}. As a result, 
 triangular and kagom\'e lattice antiferromagnets can exhibit plateaus at 1/3 of the total magnetization $m_s$ \cite{Lee84,Kawamura85, Chubukov91,Zhitomirsky02,Schulenburg02},
while the pyrochlore  and $J_1$--$J_2$ square lattice antiferromagnets both show 1/2 magnetization plateaus
\cite{Penc04,Zhitomirsky00}. 
The lattice topology, edge- versus corner-shared units, affects the periodicity and degeneracy of  emergent states and can lead to additional quantum plateaus known as  magnon crystals \cite{Schulenburg02,Zhitomirsky07,Nishimoto13,Capponi13,Picot16}.

\begin{figure}[tb]
\hskip -4mm
\includegraphics[width=0.85\columnwidth]{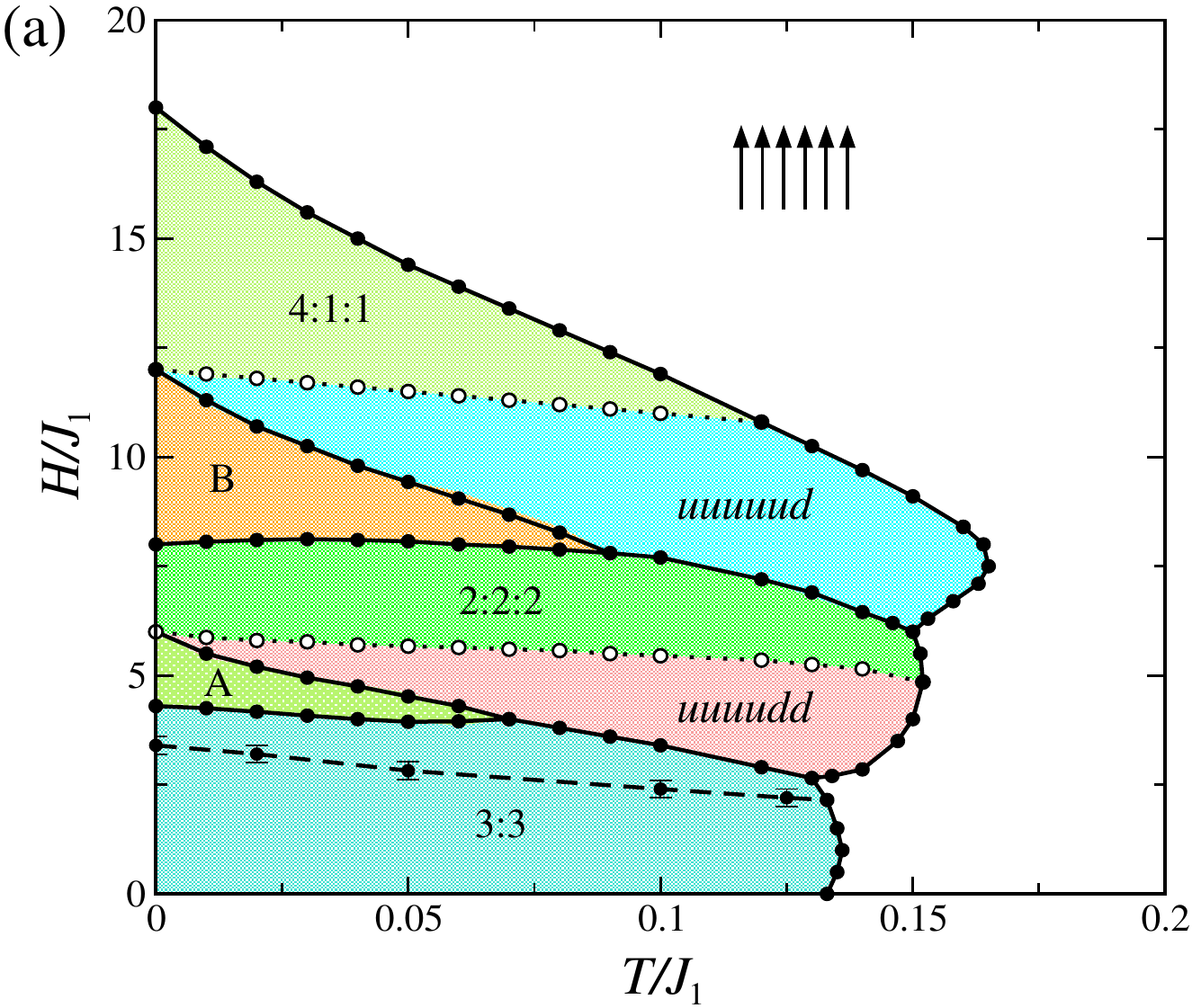}
\vskip 3mm
\centerline{
\includegraphics[width=0.88\columnwidth]{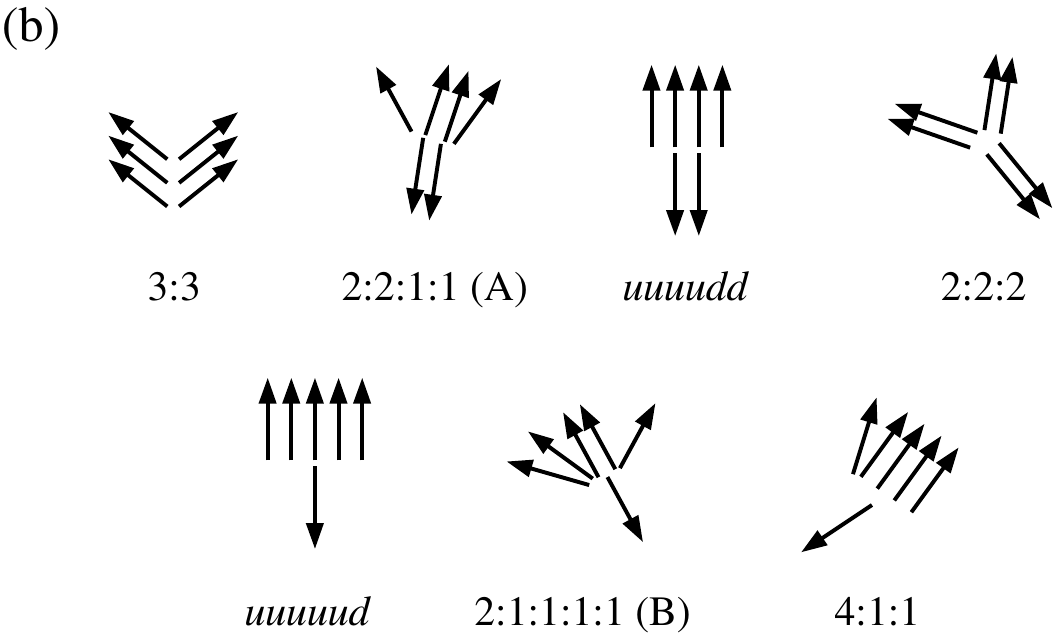}}
\caption{(a) Phase diagram of the classical Heisenberg fcc antiferromagnet with $J_2/J_1 = 1/2$. 
(b) Octahedron spin configurations for each of the phases.
Closed and open circles indicate  first- and the second-order transition points, respectively, as a guide to the eye. The dashed line and symbols  inside the 3:3 phase  indicate the first-order transitions with a change of the ordering wavevector from $\bm{Q}_3$ at low fields to  $\bm{Q}_2$ at higher fields.}
    \label{fig:HT}
\end{figure}

In this Letter, we investigate the magnetization process for the Heisenberg  
antiferromagentic model for both types of octahedral lattices:
\begin{equation}
\hat{\cal H} = 
\sum_{\left\langle ij \right\rangle} J_{ij} \, \bm{S}_i\cdot\bm{S}_j - 
\bm{H}\cdot \! \sum_{i}\bm{S}_i\,.
\label{H}
\end{equation}
Two kinds of exchange constants naturally appear for octahedral blocks: $J_1$ along the edges 
and $J_2$ connecting the opposite vertices, see Fig.~\ref{fig:geometry}(c). The magnetization plateaus 
appear at  $m/m_s=1/3$ and $2/3$. Surprisingly, we also find an abundance of various
coplanar states in between, which produce 
intricate and complex phase diagrams. An example of such a diagram for the classical  spins on the edge-shared 
octahedra network is shown in Fig.~\ref{fig:HT}.

\textit{Classical ground states.}\ --- The lowest-energy magnetic states of the spin Hamiltonian (\ref{H}) are 
obtained by treating $ \bm{S}_i$ as classical vectors of unit length. For a corner-shared lattice, 
the energy per single octahedron  can be written as 
\begin{equation}
E_{\rm oct} =\frac{1}{2}\bigl[
J_1\bm{S}^2_{\rm tot}+
(J_2-J_1)(\bm{S}_{d1}^2 + \bm{S}_{d2}^2 + \bm{S}_{d3}^2)
- \bm{H}\cdot\bm{S}_{\rm tot}\bigr],
\label{Eoct1}
\end{equation}
where $\bm{S}_{\rm tot}$ is the total spin of an octahedron unit and
 $\bm{S}_{d_n}$ are sums of spins on diagonal bonds.
For $J_2<J_1$, pairs of diagonal spins remain parallel for all $H$, maximizing $|\bm{S}_{d_n}| = 2$. In zero field,
the condition $\bm{S}_{\rm tot}=0$ leads to a $120^\circ$ arrangement of three vectors $\bm{S}_{d_n}$.
The lowest energy state of a single unit also yields a ground state of the entire system  by simple periodic
repetition, Fig.~\ref{fig:geometry}(a). The $q=0$ triangular antiferromagnetic structure has been observed in 
several cubic materials consisting of octahedra \cite{Kren68,Tomeno99,Bertaut68}. Additional ground states are constructed from the $q=0$ structure by rotating two sublattices around the third one. Such rotations can be performed independently by different angles for spins in parallel square planes sandwiched between layers of the third sublattice, see Fig.~\ref{fig:geometry}(a).
This continuous degeneracy is present for all $J_2< J_1$ and is consistent with degenerate lines
$(0,0,q)$ in the momentum space for the  lowest eigenvalue of the Fourier transform of the
exchange matrix \cite{SM}.

In an applied field, minimization of (\ref{Eoct1}) yields  the total magnetization of an octahedron unit
\begin{equation}
\bm{S}_{\rm tot} = \bm{H}/(2J_1).
\label{Stot1}
\end{equation}
This condition can be satisfied below the saturation field $H_s=12J_1$ by a family
of three-sublattice structures described by two continuous angle parameters.
Similar to the case of  a triangular-lattice antiferromagnet,
quantum or thermal fluctuations favor  the coplanar spin configurations via the order by disorder effect
\cite{Lee84,Kawamura85,Chubukov91}. In particular, they stabilize the intermediate collinear up-up-down ($uud$) state
producing the 1/3 magnetization plateau for all $J_2<J_1$. Thus, the experimental observation of such a plateau i
n Eu$_3$SnO \cite{Li23} is fully consistent with the antiferromagnetic nearest-neighbor interactions between Eu ions  in the corner-shared octahedral lattice.

The classical degeneracy 
is greatly enhanced at the fully frustrated point $J_2 = J_1$.  For equal couplings,  any six-sublattice configuration 
satisfying (\ref{Stot1}) yields the ground state for all fields $0\leq H\leq H_s$.  Furthermore,
the Maxwellian counting arguments 
\cite{Moessner98} suggest a macroscopic degeneracy of the ground-state manifold for a lattice of 
corner-shared octahedra parameterized by $N$ angles, where $N$ is the total number of spins. 
Such a degeneracy corresponds to a flat band of zero energy 
modes in the momentum space  \cite{SM} similar to kagome and pyrochlore antiferromagnets \cite{Harris92,Moessner98}. This may lead to new interesting spin liquid states  
in zero field \cite{Benton21,Paddison24}.

The ordered states of the $J_1$--$J_2$ fcc antiferromagnet include the AF2 phase for $J_2/J_1>1/2$ described 
by the propagation wavevectors $ \bm{Q}_2=(\pi,\pi,\pi)$  and  the AF3 phase for
$0<J_2/J_1<1/2$ with $\bm{Q}_3=(0,\pi,2\pi)$
\cite{Seehra88}. 
The degeneracy is enhanced at the boundary $J_2/J_1=1/2$, 
where the classical ground states are spin spirals satisfying
\begin{equation}
\cos\frac{Q_x}{2} + \cos\frac{Q_y}{2} + \cos\frac{Q_z}{2} = 0\,,
\label{Q}
\end{equation}
 see  Fig.~\ref{fig:geometry}(d) \cite{Lines64,Heinila93,Balla20}. 
At this fully frustrated point, the block representation applies to  the fcc model (\ref{H})  
with an octahedron energy
\begin{equation}
E_{\rm oct} = \frac{1}{4}\, J_1 \bm{S}^2_{\rm tot} - \frac{1}{6}\,\bm{H}\cdot \bm{S}_{\rm tot} \,.
\label{Eoct2}
\end{equation}
The difference in the prefactors from Eq.~({\ref{Eoct1}) is due to edge-shared versus corner-shared stacking of octahedra.
The lowest energy of a single octahedron block is obtained for
\begin{equation}
\bm{S}_{\rm tot} =  \bm{H}/(3J_1)\,.
\label{Stot2}
\end{equation}
Just as for the corner-shared lattice, any six-sublattice configuration with a fixed magnetization ({\ref{Stot2}) gives 
a classical ground state.  Thus, two fully-frustrated octahedral models exhibit similar local degeneracy.
In a magnetic field they can exhibit a rich variety of ordered magnetic structures
stabilized by the effect of order by disorder \cite{Villain80,Shender82}.
The block connectivity, edge vs.\ corner-shared, determines the ordering wavevectors and the 
additional degeneracy that may arise once six sublattices are distributed over the entire lattice.

\textit{Single octahedron states.}---We begin by exploring possible magnetic states for a fully frustrated
octahedron unit. Six-sublattice configurations satisfying (\ref{Stot1}) (or (\ref{Stot2})) are parameterized 
by eight continuous variables, modulo a global rotation about  $\bm{H}$. 
This leads to a significantly larger manifold of classical ground states compared to spin models based on 
triangles (two angles) or tetrahedra (four angles).  Consequently, the selection of states by quantum 
or thermal fluctuations becomes a nontrivial problem.

In general, short-range quantum fluctuations can be taken into account by
an effective biquadratic exchange \cite{Henley89,Heinila93a,Zhitomirsky15}. Accordingly, we 
consider
\begin{equation}
\hat{\cal H}_{\rm biq} = - \sum_{\langle ij \rangle} B_{ij} ( \bm{S}_i \cdot  \bm{S}_j)^2 
\label{Hbiq}
\end{equation}
with two dimensionless constants $b_{1,2} = B_{ij}S^2/J_{ij}$ for the  first ($J_1$) and 
the second ($J_2$) neighbor bonds.
The second-order quantum correction to the ground state energy yields 
$b_1 = 1/(24S)$, $b_2/b_1=1/2$ for the fully-frustrated fcc model and
$b_1 = 1/(16S)$, $b_2/b_1=1$ for the corner-shared lattice with $J_2=J_1$, 
see \cite{SM}.  For purely classical spin models,  the  low-$T$ contribution of the short-range
thermal fluctuations can be also expressed  in a biquadratic form with $b\propto T/J$  \cite{Canals04}. In addition,
magnetoelastic interactions \cite{Penc04} or itinerant electrons \cite{Wysocki13} may contribute to an
effective  biquadratic interaction in a real material.

The equilibrium spin configurations for a single octahedron
are found by numerical minimization of  (\ref{Eoct2}) with extra biquadratic interactions
(\ref{Hbiq}). Results are presented in the $H$--$b$ phase diagram of Fig.~\ref{fig:Hb}.
The ratio of biquadratic constants was fixed to 1/2 obtained above for
the fully-frustrated fcc model.
Generally, a negative biquadratic interaction (\ref{Hbiq}) favors collinear structures
among degenerate classical states.
In particular, a collinear state is found in zero field, which transforms into a canted 3\,:\,3 structure
for small $H$.
A frustrated octahedron unit exhibits two additional collinear states at intermediate fields: 
up-up-up-up-down-down ($uuuudd$) and $uuuuud$ that are stable at
 $H=H_s/3$ and $H=2H_s/3$, respectively.  Biquadratic interactions  
 transform the two isolated points into finite-width magnetization plateaus.

\begin{figure}[tb]
\centering
\includegraphics[width=0.99\columnwidth]{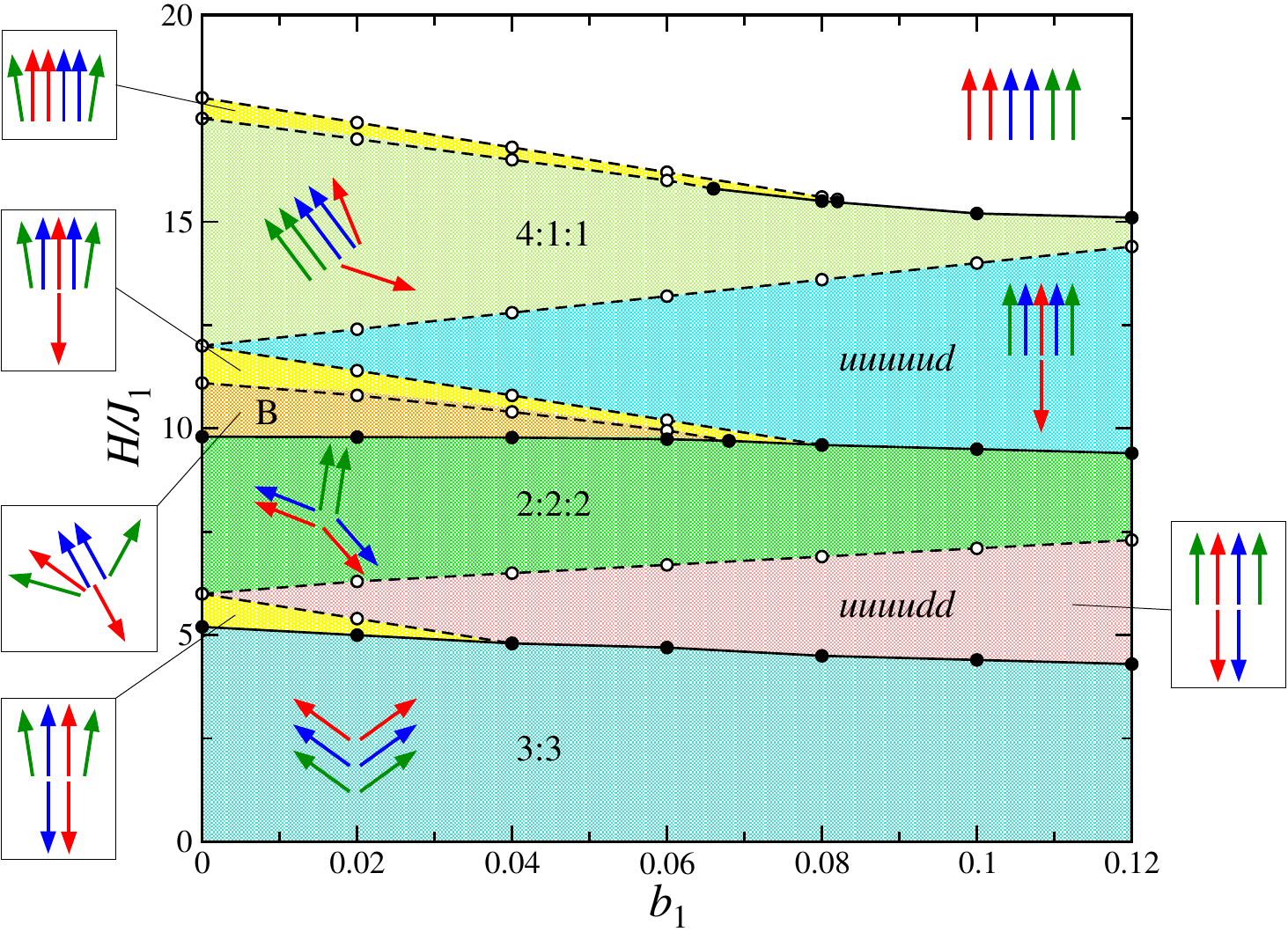}
\caption{Magnetic structures of a spin octahedron with a fixed ratio of biquadratic exchange
constants $b_2/b_1=1/2$. 
Closed circles and solid lines indicate first-order transitions, open circles and dashed lines denote
second-order transitions. Arrows with the same color represent spins on opposite vertices of an octahedron.}
\label{fig:Hb}
\end{figure}

The phase diagram includes  a variety of coplanar magnetic states.
Their stability and occupied regions are strongly 
influenced by biquadratic terms. The largest number 
of coplanar states is present for small $b_1$  suggesting that semiclassical, large-$S$ antiferromagnets  generally have
more complex phase diagrams compared to the quantum models with  $S=1/2$ 
($b_1=0.083$) or 1 ($b_1=0.041$). For sufficiently large $b_1\agt 0.075$, only three of these
coplanar structures survive, 3\,:\,3, 2\,:\,2\,:\,2, and 4\,:\,1\,:\,1, all exhibiting a partial collinearity
between magnetic sublattices.  These states, as well as the collinear $uuuudd$ and  $uuuuud$ structures,
break the octahedron symmetry according to specific irreducible representations of the 
point group $O_h$ and hence are stable with respect to additional interactions. 
Other values of $b_2/b_1$ can also appear in real materials either
as a result of higher-order quantum effects or due to simultaneous presence of several mechanisms for
an effective biquadratic interaction. Stable spin configurations for $b_2/b_1=1/4$ and 1 
are listed in Appendix A together with further details on the $H$--$b$ diagram.

\textit{Fully-frustrated fcc model:~Monte Carlo results.}\ --- To extend the preceding analysis based 
on local spin correlations, we now investigate the full effect of 
fluctuations in a lattice by performing the Monte Carlo (MC) simulations of the classical fully-frustrated fcc model (\ref{H}). 
The standard Metropolis algorithm combined with microcanonical over-relaxation steps have been used
to simulate periodic clusters of linear sizes $L=6$--24 with $N=L^3$ spins. Further details on the algorithm implementation 
are provided in Supplemental Material \cite{SM}.

\begin{figure}[tb]
\centering
\includegraphics[width=0.85\columnwidth]{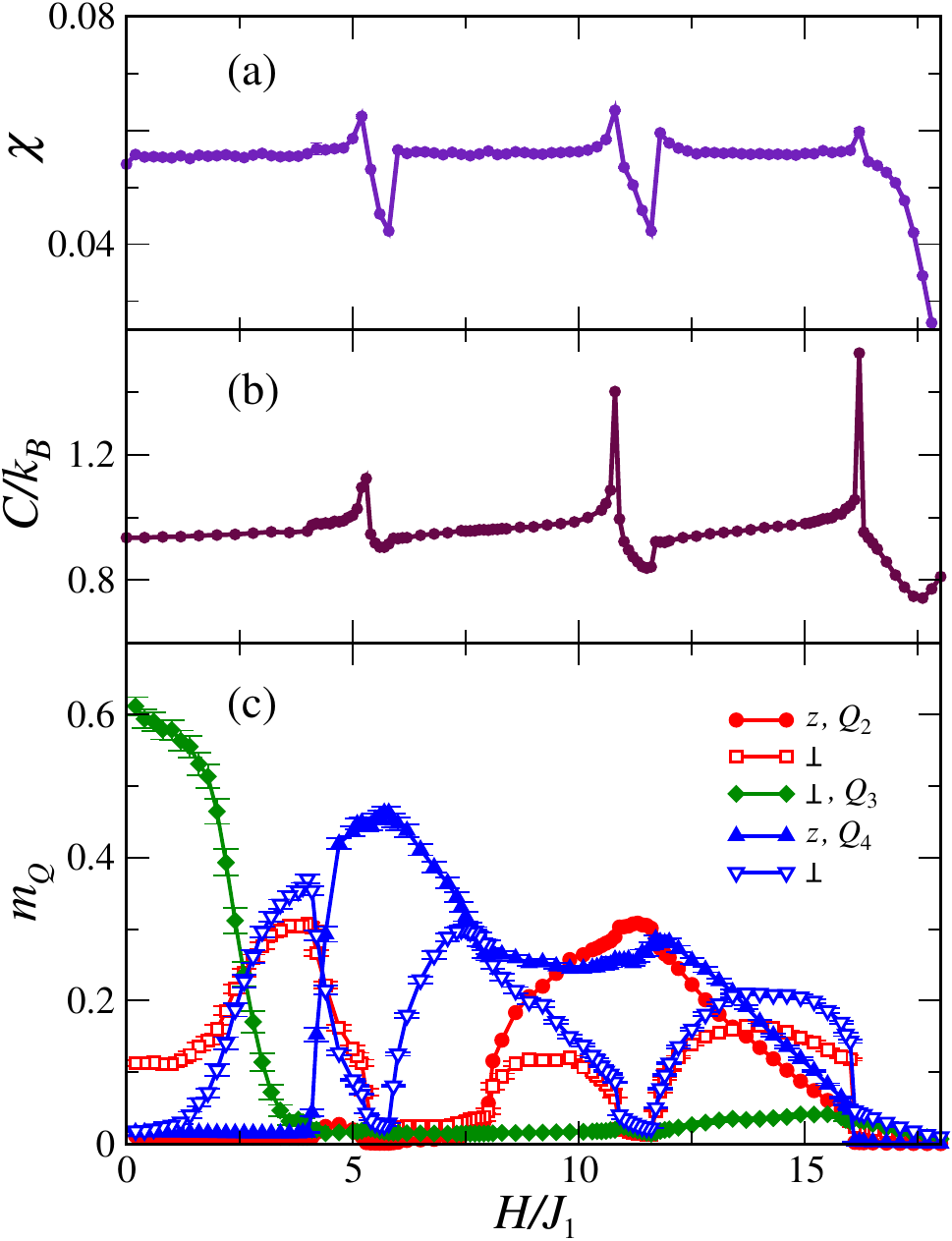}
\caption{Magnetic field scans from classical Monte Carlo simulations of the fully-frustrated fcc antiferromagnet ($J_2=J_1/2$)  at $T/J_1=0.02$: (a) differential magnetic susceptibility $\chi= dM/dH$,
(b) specific heat  (both per spin), (c) longitudinal  $|m_Q^z|$ ($z$)  and transverse $m^\perp_Q = 
(m_Q^{x2} + m_Q^{y2} )^{1/2}$  ($\perp$) components of the Fourier harmonics 
of the antiferromagnetic order parameter, see caption pf Fig.~1 and text for definitions of $Q_n$. 
The Monte Carlo results have been obtained for the $L=24$ cluster.
}
\label{fig:MC}
\end{figure}

Representative field scans for several physical quantities  are shown in Fig.~\ref{fig:MC} for  $T/J_1=0.02$. 
Anomalies in the differential magnetic susceptibility $\chi = dM/dH$ and the specific heat $C$ 
have been used to locate the  phase transitions. In particular,  two dips in $\chi$ at  $H_{c1} \approx 6J_1$ 
and  $H_{c2} \approx 12J_1$ indicate  the 1/3 and 2/3 magnetization plateaus. 
In classical frustrated spin models, 
$\chi$ typically remains finite in the  plateau region instead of dropping to zero \cite{Kawamura85,Zhitomirsky02}.
Such a behavior is related to absence of a quantum gap and, hence, to a large
density of gapless excitations at finite $T$. Presence of the
collinear $uuuudd$ and $uuuuud$ states is clearly supported 
by the magnetic order parameters discussed next.

The Fourier harmonics of the antiferromagnetic order parameter
 \begin{equation}
 m_{\bm{q}}^\alpha = \frac{1}{N} \sum_i S_i^\alpha\, e^{-i \bm{q}\bm{r}_i}\,.
 \label{mQ}
 \end{equation}
 have been measured during the MC simulations.
Two relevant modes are $\bm{Q}_2$ and $\bm{Q}_3$,  which correspond to 
zero-field states of the $J_1$--$J_2$ fcc antiferromagnet. Other wavevectors from 
the soft mode surface  (\ref{Q}) may also contribute. In the auxiliary MC runs on small
clusters ($L=6$--14), we have identified another relevant harmonic corresponding to the $G$ point with  
$\bm{Q}_4=(4\pi/3, 0, 4\pi/3)$, see Fig.~\ref{fig:geometry}(d). 
This point lies at the cross-section 
of the surface  (\ref{Q}) with two 
mirror planes and, hence, corresponds to a high symmetry position similar to the $L$ and $W$ points. 
Due to the  simultaneous importance of $\bm{Q}_3$ and $\bm{Q}_4$ modes, 
which possess the four- and the three-fold periodicity, 
we performed extensive MC simulations for two lattice sizes
$L=12$ and 24. 

Figure~\ref{fig:MC}(c)  shows the  field dependence of antiferromagnetic  modes (\ref{mQ}),
which have been  summed over the star $\{\bm{Q}_n\}$ of each wavevector.
In the high field region $H/J_1 \agt 3$, the equilibrium magnetic states 
are composed of the harmonics that belong either to   $\{\bm{Q}_2\}$  or to  $\{\bm{Q}_4\}$.
At the magnetization plateaus, the transverse components
of the antiferromagnetic order parameter  $m^\perp_Q = (m_Q^{x2} + m_Q^{y2} )^{1/2}$ vanish, whereas
the longitudinal modes  $|m^z_Q|$ stay finite confirming the collinear spin configuration. The additional MC data  
give $\langle |S_i^z|\rangle\approx 0.9$ at $T/J_1=0.02$, which demonstrates a fully developed longitudinal order for both magnetization plateaus \cite{SM}.

The classical degeneracy for spin spirals with different wavevectors belonging to the surface (\ref{Q}) 
holds for all fields $H<H_s$. Nevertheless,  only a few antiferromagnetic harmonics  appear in the Monte Carlo simulations. This is a clear indication of the thermal order by disorder selection produced by the entropic term in the free energy.
Specifically,  the $uuuudd$ state at the 1/3  plateau is a single-$k$ structure
based  on one of the $\bm{Q}_4$ harmonics. The 
$uuuuud$ magnetic structure at the 2/3  plateau  is a double-$k$ state that mixes
 the $\bm{Q}_2$ and  $\bm{Q}_4$ modes. More details on the real space magnetic structures for the two plateau
 states are provided in Appendix B.

The $H$--$T$ diagram shown in Fig.~\ref{fig:HT} combines the MC results from multiple field and 
temperature scans. Its eight antiferromagnetic phases are labeled according to spin configurations 
that are the same for all  octahedral blocks in the lattice. To perform such an identification, we have used 
an octupole order parameter $T^{\alpha\beta\gamma} = \langle S_i^\alpha S_i^\beta S_i^\gamma\rangle$
($\alpha,\beta,\gamma = x,y$) as well as instantaneous spin configurations, 
see \cite{SM}.  The first-order transition lines  are shown by full lines with black circles, 
whereas second-order transitions are indicated by dotted lines and open symbols. The dashed line 
inside the region of 3\,:\,3 phase corresponds to a first-order transition with
a change of the ordering wavevector from $\bm{Q}_2$, at higher fields,  to
$\bm{Q}_3$, at lower fields. Such a transition is apparent in the field dependence of  
corresponding $m^\perp_Q$ in Fig.~\ref{fig:MC}(c).

The low-$T$ part of the Monte Carlo phase diagram, Fig.~\ref{fig:HT}, has an apparent similarity to the 
$H$--$b$ diagram of the single-octahedron model, see Fig.~\ref{fig:Hb}. 
This is because a low-$T$ contribution from thermal fluctuations can be expressed as an effective 
$T$-dependent biquadratic coupling \cite{Canals04}. Nonetheless, there are a few notable differences.  For example,
the $C_2$ symmetric coplanar structures present in Fig.~\ref{fig:Hb} disappear completely 
in the MC diagram. An opposite trend affects the low-symmetry phase A, which occupies a finite region 
in the $H$--$T$ diagram, while its presence in the mean-field $H$--$b$ diagram is restricted to a tiny space  
not visible  on the scale of  Fig.~\ref{fig:Hb}, see \cite{SM}.

\textit{Discussion.}\ --- The $J_1$--$J_2$ Heisenberg fcc antiferromagnet with $J_2/J_1 = 1/2$
is an example of the fully frustrated spin model on the edge-shared octahedral lattice.
The classical Monte Carlo simulations reveal its rich and complex behavior in external field 
with eight different antiferromagnetic phases appearing below the saturation field. 
A distinctive feature of the magnetic frustration in octahedron blocks is the presence of two stable magnetization 
plateaus at fractional values of $m=1/3$ and $m=2/3$.
For a better understanding of emergent magnetic structures, we also studied a toy model of a single spin octahedron  with additional effective biquadratic exchange. The calculated phase diagram of the toy model  provides 
an accurate guide to low temperature phases obtained in the MC study of the lattice spin model.

The fully frustrated spin octahedra  can exist in the form of a corner-shared lattice 
in magnetic antiperovskites. In particular, the magnetization curves of Eu$_3$PbO and  
Eu$_3$SnO  exhibit several plateau-like features \cite{Hirschmann22,Li23}.
Further experimental work is needed to clarify their relation with magnetic frustration in the octahedral units.
On the theoretical side, the  $J_1$--$J_2$ classical spin model is highly degenerate for the corner-shared
lattice of octahedra.  Strong magnetic frustration for $J_2=J_1$
promotes the classical spin liquid states
 \cite{Benton21,Szabo22,Paddison24} and precludes the thermal order by disorder effect. 
However, similar to the previous work on the pyrochlore antiferromagnet
\cite{Shannon10}, the further-neighbor
exchanges can stabilize the $q=0$ magnetic structures making a single-octahedron physics relevant again.

Finally, the fully frustrated six-spin units that are similar, but not identical to the octahedron blocks
discussed here, also arise for the $J_1$--$J_2$--$J_3$ antiferromagnet on a honeycomb lattice for a special 
relation between three exchange constants \cite{Rehn16,Albarracin21}.
Overall, our results illustrate new interesting aspects of magnetic frustration in octahedral lattices.

\textit{Acknowledgments.}\ --- We thank M. Gingras, H. Takagi, and A. Zelenskiy for useful discussions. 
The work of ASG was  financially supported  by the French Research Agency (ANR) within the Project
Fresco, No.\ ANR-20-CE30-0020. MEZ acknowledges a partial support by the ANR Project
Fragment,  Project No.\  ANR-19-CE30-0040.



\newpage
\onecolumngrid
\begin{center}
{\large\bf End Matter}
\end{center}
\vspace*{5mm}

\twocolumngrid
\renewcommand{\theequation}{A\arabic{equation}}
\setcounter{equation}{0}

{\it Appendix A: Single octahedron states.}\ --- 
To demonstrate the stability and ubiquity of the coplanar magnetic states described in the main text,
we present here additional numerical results for other values of  $b_2/b_1$. 
Figure~\ref{Fig5} shows the magnetization curves $M(H)$ with $dM/dH$ (upper panel)
and field-induced phases (lower panel)
for $b_1 = 0.02$ and $b_2/b_1 = 1/4, 1/2$, and 1. The magnetization curves are very similar, each exhibiting
two magnetization plateaus of comparable width. 
The case $b_2=b_1$ has an additional symmetry and its states are discussed later.
On the whole, there is a remarkable correspondence between stable spin structures obtained  for all $b_2<b_1$.
The six states,  4\,:\,1\,:\,1, $uuuuud$,  
2\,:\,1\,:\,1\,:\,1, 2\,:\,2\,:\,2,   $uuuudd$,  and 3\,:\,3 (from top to bottom), are common for $b_2/b_1 = 1/4$ and 1/2.
The rest of the differences can be understood from the following analysis.

\begin{figure}[t]
\centerline{
\includegraphics[width=0.83\columnwidth]{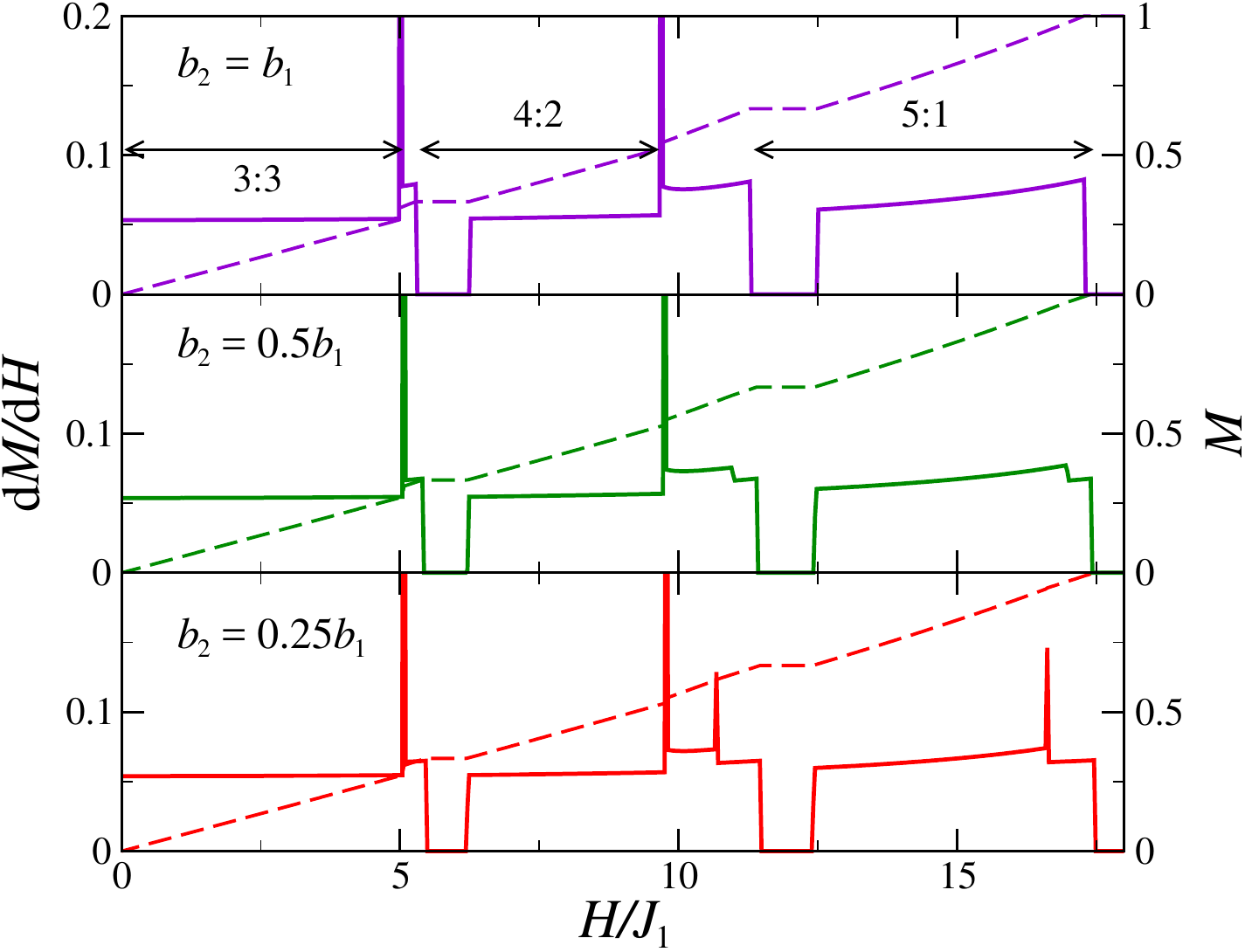}}
\vskip 2mm
\centerline{
\includegraphics[width=0.85\columnwidth]{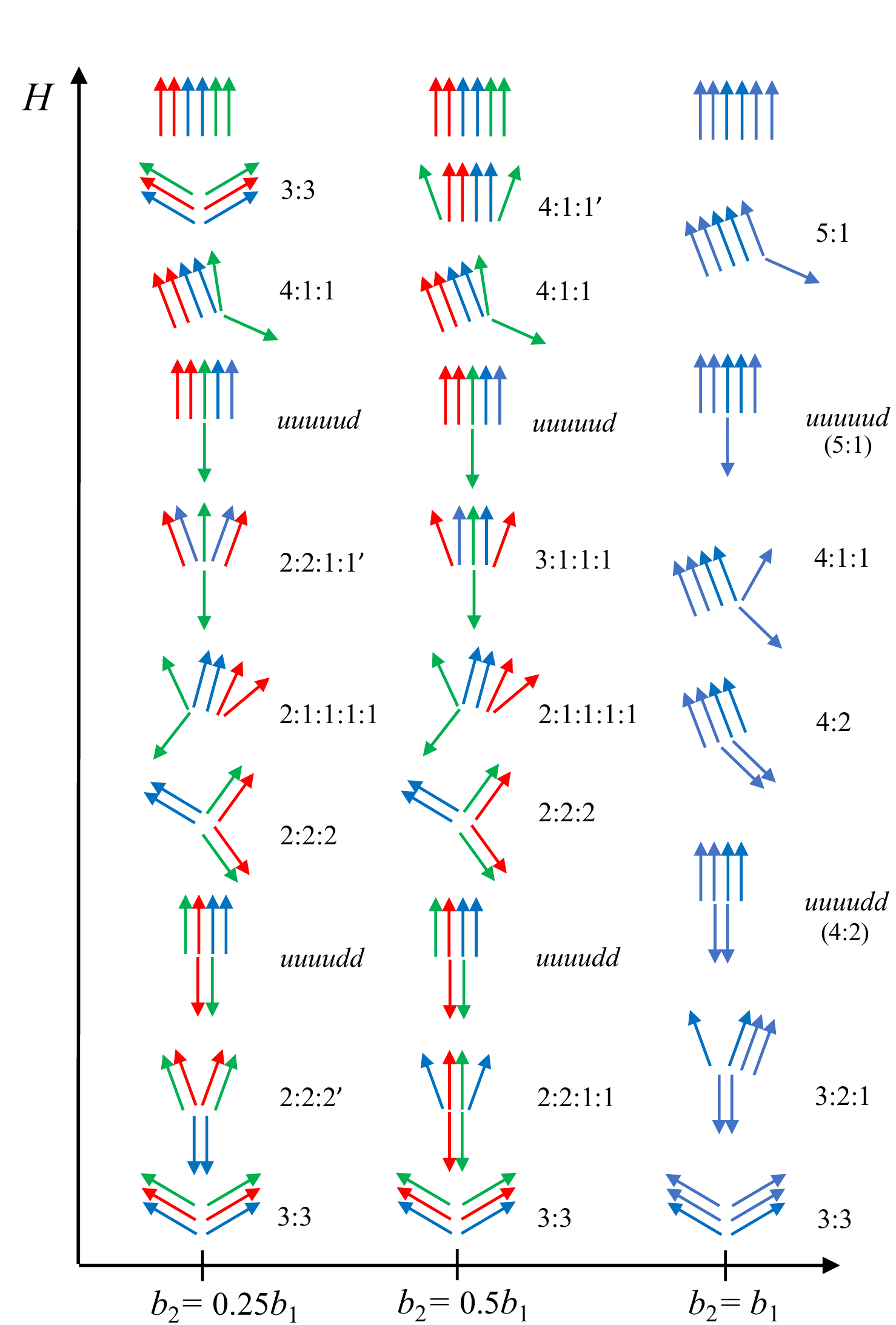}}
 \caption{
 Upper panel: the magnetization curves $M(H)$ (dashed lines) and the differential susceptibility $dM/dH$ 
 (solid lines) versus  field for $b_1 = 0.02$ and varying $b_2/b_1$.
 Lower panel:
Graphical representation of stable spin configurations in various fields. 
Vectors of the same color  for $b_2/b_1<1$  indicate spins on the opposite vertices of an octahedron.
}
\label{Fig5}
\end{figure}

The symmetry group of the isotropic model in a magnetic field is ${\cal G}= O_h\times U(1)$ with
uncoupled spin and space rotations. The fully polarized  $uuuuuu$ state 
is invariant under all symmetry operations of $\cal G$.  The second-order transition field 
$H_s$ into a canted antiferromagnetic state is determined analytically by expanding 
the energy in transverse  spin components and diagonalizing the obtained quadratic form:
\begin{equation}
  H_s = 6J_1(3 - 8b_1 - 4b_2) \,.
\end{equation}
The corresponding eigenmodes transform according to the $T_{1u}$  irrep   
of $O_h$. The symmetry of the antiferromagnetic state 
is established  by considering quartic terms in the energy expansion. 
Analytically, an effective `cubic' anisotropy in the Landau functional
for the $T_{1u}$ mode changes sign at $b_2=0.4b_1$. For $b_2 < 0.4b_1$,
the three  $T_{1u}$ components have equal amplitudes producing
the 3\,:\,3 canted structure. For $b_2 > 0.4b_1$, only 
one component of the $T_{1u}$ triplet is selected, resulting in the 4\,:\,1\,:\,1'  state, see Fig.~\ref{Fig5}.

A similar analysis can be performed  for each of the magnetization plateaus. 
The $uuuuud$ state  of the 2/3 plateau preserves  ${\cal G}'= C_{4v}\times U(1)$ symmetry.
The upper and lower plateau boundaries are given, respectively, by  
\begin{eqnarray}
    H_{c_2}^{(2/3)} & = &12J_1\bigl(1  + \sqrt{17b_1^2 + 8b_1b_2} - b_1\bigr),  \nonumber \\
    H_{c_1}^{(2/3)}  & = & 12J_1(1 - 4b_1 - 2b_2)\,.
\end{eqnarray}
The instability at the upper edge of the plateau is determined by a single mode that transforms 
as $A_1$ irrep of $C_{4v}$. At the lower edge, a two-component mode of the $E$ irrep condenses.
Again, we find the mixed two-component $E$ state  2\,:\,2\,:\,1\,:\,1' for $b_2< 4b_1/9$,
whereas the single-component $E$ state 3\,:\,1\,:\,1\,:\,1  is energetically  stable for $b_2> 4b_1/9$.
Similarly, for the 1/3 magnetization plateau we find
\begin{eqnarray}
    H_{c_2}^{(1/3)} & = & 6J_1\bigl[1 + 4\sqrt{b_1(3b_1+b_2)} - 4b_1\bigr],         \nonumber \\
    H_{c_1}^{(1/3)}  & = & 6J_1(1 - 8b_1 - 4b_2)\,.
\end{eqnarray}

Note, that the $C_2$ symmetric coplanar phases that exist below the collinear states
 in the mean-field theory are absent in the MC simulations. 
Instead, the lattice model exhibits  direct first-order transitions into the less symmetric structures.

The symmetry of a single spin octahedron is enhanced for equal biquadratic constants $b_2=b_1$.
There is no longer a distinction between edge and diagonal bonds of an octahedron. 
Accordingly, the block Hamiltonian is symmetric  with respect to  all permutations of six spins, which
form the group $S_6$.
Stable magnetic structures obtained in this case are shown in Fig.~\ref{Fig5}.
A partial collinearity in these canted spin structures becomes even more apparent. In fact,
magnetic states found for  $b_2=b_1$ belong 
 to different irreducible representations of $S_6$ that  are described by the Young diagrams.  
 This is emphasized by the chosen notations for the coplanar spin configurations, 
 which list the number of boxes in each row of the corresponding Young diagram.   
Interestingly, the high-field 4\,:\,1\,:\,1  state present for $b_2<b_1$
is obtained by an appropriate spin distortion from the 5\,:\,1 state found for $b_2=b_1$.
Such a relation explains why the  low-symmetry 4\,:\,1\,:\,1  state has a wide range of stability
for different $b_1$ and $b_2$.  At $b_2=b_1$, the crossing of different irreps of $O_h$ takes place. 
Accordingly, modifications of field-induced states occur for $b_2>b_1$. 
 At present it is not clear whether such a combination of effective biquadratic parameters can be realized
 in magnetic materials.
 We leave the necessary analysis of this case for future work.

\begin{figure}[tb]
\centerline{
\includegraphics[width=0.75\columnwidth]{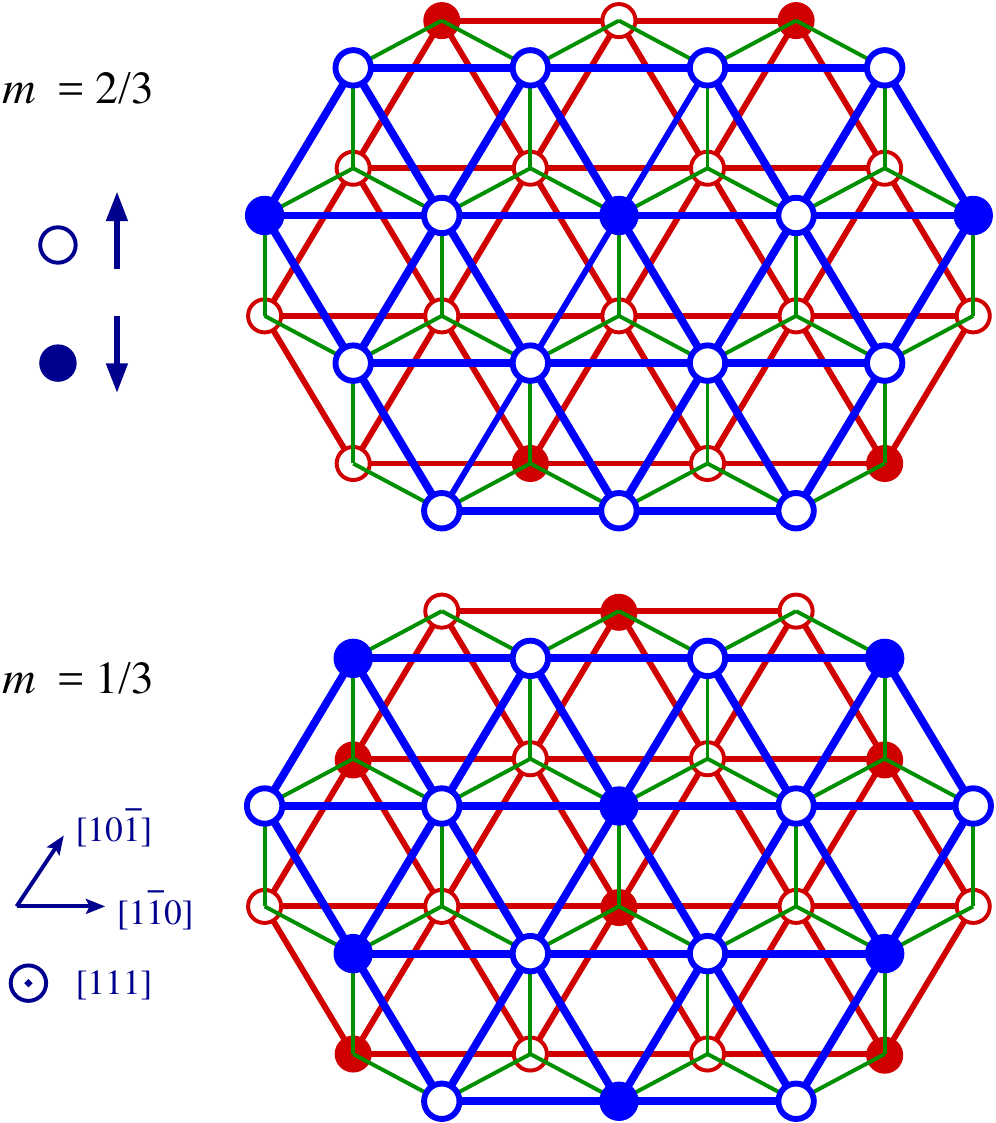}
}
 \caption{The high symmetry magnetic structures for the $m=1/3$ (bottom panel) and the $m=2/3$ (top panel) 
 magnetization plateaus. The fcc lattice is 
 represented as an $ABC$ stack of triangular layers of magnetic ions. Two such adjacent layers 
 are shown by blue (upper layer) and dark  red (lower layer) circles. 
Octahedron spin blocks  formed between the two layers 
are indicated by green hexagons.  
 }
\label{Fig6}
\end{figure}

{\it Appendix B:\ Mangetization plateau structures.}\ ---
Here we describe magnetic structures for the two magnetization plateaus found in the MC simulations  
of the fully-frustrated fcc model.  It is convenient to represent a fcc lattice by the $ABC$ stack of 
triangular planes of magnetic atoms along the [111] crystal axis. Figure~\ref{Fig6} shows two 
adjacent triangular layers drown by thick (top layer) and thin (bottom layer) lines. 
An octahedral block in this two-dimensional projection is represented by 
overlapping $\bigtriangleup$ and $\bigtriangledown$ triangles from adjacent layers with interlayer $J_1$ 
bonds forming the hexagons. The up ($\uparrow$) and down ($\downarrow$) spins are shown by open and 
filled circles.

The high-symmetry single-$k$ state realized at the 1/3 magnetization plateau in the Monte Carlo simulations 
is shown in the bottom panel of  Fig.~\ref{Fig6}. Every triangular layer has a $\sqrt{3}\times\sqrt{3}$ superstructure 
of $\downarrow$ spins. The classical constraint (\ref{Stot2}) applied to  the $uuuudd$ state requires two  
$\downarrow$ spins per every octahedron. This constraint can be straightforwardly verified for the presented
spin configuration. Let us choose a $\downarrow$ spin in the top layer.  
The constraint for the two adjacent octahedra is satisfied if a $\downarrow$ spin in the bottom layer is placed 
on one of the three vertices of the opposite triangular plaquette. 
The same $\sqrt{3}\times\sqrt{3}$  structure of $\downarrow$ spins  
in the two layers extends 
the constraint to all octahedra sandwiched between them.
A similar step  with a choice between three possible positions for the initial $\downarrow$ spin
has to be made for the next $C$ layer  and so on. As a result, the degeneracy  of $uuuudd$ 
ground states amounts to at least $3^{L_p}$, with $L_p$ being the number of 
triangular layers. A state with the smallest magnetic unit cell  is constructed if 
interlayer bonds  with two $\downarrow$ spins  are all parallel to the same nearest-neighbor direction.
One such state corresponding to $\bm{Q}_4 = (4\pi/3,-4\pi/3,0)$ is shown in Fig.~\ref{Fig6}.
 
For the upper magnetization plateau,  the density of down spins in every triangular layer must be half
that of the 1/3 plateau state. This can be achieved by placing alternating $\uparrow\downarrow$ spins 
on every third chain in a triangular layer, while two intermediate spin chains remain fully polarized, as shown in 
the top panel of Fig.~\ref{Fig6}. In performing such a coloring one must select for every third chain one of 
two $\uparrow\downarrow\uparrow\downarrow$ structures. This yields the
$2^{L_c}$ configuration degeneracy; $L_c$ being a number of parallel chains in one layer. Once a choice 
of $\downarrow$ spins is made for the entire $A$ layer, the classical constraint  determines
uniquely the $\downarrow$ spin pattern in the adjacent $B$ layer and so on. 
The double-$k$ magnetic structure with a unit cell of six spins shown in Fig.~\ref{Fig6} is constructed as
a superposition of the Fourier harmonics with $\bm{Q}_2 = (\pi,-\pi,\pi)$ and $\bm{Q}_4 = (4\pi/3,4\pi/3,0)$,
which is consistent with the MC results presented in  Fig.~\ref{fig:MC}(c).

\newpage

\onecolumngrid
\begin{center}
{\large\bf 
Fully-frustrated octahedral antiferromagnets: emergent complexity in external field \\[1.5mm]
Supplemental Material } \\
\vskip 3.5mm
A. S. Gubina,$^1$, T. Ziman,$^{2,3}$, and M. E. Zhitomirsky$^{1,2}$ \\
\vskip 1.5mm
{\it \small $^1$Universit\'e Grenoble Alpes, CEA, IRIG, PHELIQS, 38000 Grenoble, France} \\
{\it \small $^2$Institut Laue Langevin, 38042 Grenoble Cedex 9, France} \\ 
{\it \small $^3$Kavli Institute for Theoretical Science, University of the Chinese Academy of Sciences, 100190, Beijing,  China}
\\[0.5mm]
{\small (Dated: February 4, 2025)}
\end{center}
\vspace*{5mm}

\setcounter{page}{1}
\thispagestyle{empty}
\makeatletter
\renewcommand{\c@secnumdepth}{0}
\makeatother
\setcounter{section}{0}
\renewcommand{\theequation}{S\arabic{equation}}
\setcounter{equation}{0}
\renewcommand{\thefigure}{S\arabic{figure}}
\setcounter{figure}{0}

\section{Corner-shared octahedral model: degeneracy in momentum space}

The lattice of corner-shared octahedra in cubic antipervoskites has a unit cell with three 
magnetic ions located at
\begin{equation}
\bm{\rho}_1 = (1/2, 0, 0)\,, \quad \bm{\rho}_2 = (0, 1/2, 0)\,, \quad \bm{\rho}_3 = (0, 0, 1/2)\,. 
\end{equation}
Accordingly, each spin is described by two indices:  $\bm{S}_{ni}$, where $i$ denotes a unit cell position on 
the cubic lattice and $n=1,2,3$ is a sublattice index inside the unit cell.

We consider the Heisenberg spin model on a corner-shared octahedral lattice with $J_1$
exchange constant for the nearest neighbors and $J_2$ exchange between the second-neighbor spins
{\it inside} octahedra. Explicitly, this spin Hamiltonian can be written as

\begin{eqnarray}
\hat{\cal H} & =  & J_1 \sum_i \Big[  \bm{S}_{1,i} \cdot\big( \bm{S}_{2,i} + \bm{S}_{2,i+\bm{x}} + 
\bm{S}_{2,i-\bm{y}} + \bm{S}_{2,i+\bm{x}-\bm{y}}\big)  +  \bm{S}_{1,i} \cdot \big(\bm{S}_{3,i} + \bm{S}_{3,i+\bm{x}} 
+ \bm{S}_{3,i-\bm{z}} + \bm{S}_{3,i+\bm{x}-\bm{z}}\big)  \nonumber \\
&  &  \mbox{} + \bm{S}_{2,i} \cdot \big(\bm{S}_{3,i} + \bm{S}_{3,i+\bm{y}} +\bm{S}_{3,i-\bm{z}} 
+ \bm{S}_{3,i+\bm{y}-\bm{z}}\big)\Big] + J_2 \sum_i \Big[\bm{S}_{1,i} \cdot \bm{S}_{1,i+ x}  +  
    \bm{S}_{2,i}\cdot \bm{S}_{2,i+ \bm{y}}  + \bm{S}_{3,i}\cdot\bm{S}_{3,i+ z} \Big]\,. \nonumber 
\end{eqnarray}

At high temperatures, the momentum-dependent 
susceptibility is expressed as 
\begin{equation}
\chi^{-1}_{mn}({\bf q},T) = \frac{3T}{S(S+1)} + J_{mn}({\bf q}) \,,
\end{equation}
where $J_{mn}({\bf q})$ is the Fourier transform of the exchange matrix and $m,n$ are sublattice indices. 
The divergence of $\chi_{mn}({\bf q},T_c)$ signifies a phase transition, where  $T_c$ and the ordering wavevector 
determined by the lowest
eigenvalue $\lambda_0({\bf q})$ of the exchange matrix $J_{mn}({\bf q})$.

Considering spins as  classical unit length vectors and performing 
Fourier transformation, we obtain for the classical energy 
\begin{eqnarray}
E & = & J_1 \sum_q \Big[\bm{S}^*_{1,\bm{q}}\bm{S}_{2,\bm{q}}(1+e^{iq_x})(1+e^{-iq_y}) + \bm{S}^*_{1,\bm{q}}
\bm{S}_{3,\bm{q}}(1+e^{iq_x})(1+e^{-iq_z}) + \bm{S}^*_{2,\bm{q}}\bm{S}_{3,\bm{q}}(1+e^{iq_y})(1+e^{-iq_z})\Big]  \nonumber \\
   & & \mbox{}+ J_2 \sum_q \Big[(c_x |\bm{S}_{1,\bm{q}}|^2 + c_y |\bm{S}_{2,\bm{q}}|^2 + c_z |\bm{S}_{3,\bm{q}}|^2 )\Big]\,, \qquad   \textrm{with}\quad  c_\alpha = \cos{q_\alpha}\ .
\end{eqnarray}
Hence, the Fourier transform of the exchange matrix is
\begin{equation}
J_{nm}(\bm{q}) =  J_1
\begin{bmatrix}
2j_2c_x                            &  (1+e^{iq_x})(1+e^{-iq_y})   &   (1+e^{iq_x})(1+e^{-iq_z})   \\
(1+e^{-iq_x})(1+e^{iq_y}) &      2j_2 c_y                         &   (1+e^{iq_y})(1+e^{-iq_z})   \\
(1+e^{-iq_x})(1+e^{iq_z}) &  (1+e^{-iq_y})(1+e^{iq_z})   &           2j_2c_z
\end{bmatrix}, \quad j_2 = J_2/J_1\,.
\end{equation}
Its eigenvalues $\lambda$ are determined by roots of the cubic equation for $t=\lambda/(2J_1)$

\begin{eqnarray}
&& (t - j_2c_x ) (t - j_2c_y ) (t - j_2 c_z) 
-   (t-j_2c_x)(1+c_y)(1+c_z) - (t-j_2c_y)(1+c_x)(1+c_z) 
\nonumber \\
&& 
\mbox{\phantom{$(t - j_2c_x ) (t - j_2c_y )  
$}}
 - (t-j_2c_z)(1+c_x)(1+c_y)  - 2(1+c_x)(1+c_y)(1+c_z) = 0\,.
\label{cubic}
\end{eqnarray}
\vskip 1mm
\noindent
By analyzing the roots of Eq.~(\ref{cubic}) we arrive at the following behavior of the lowest eigenvalue:
\begin{eqnarray}
&& J_2 < J_1\,,\quad  \lambda_0 = -4J_1 + 2J_2\,, \quad\  \textrm{for}\quad   c_x=c_y = 1, \forall c_z  
\nonumber \\
&& J_2 = J_1\,,\quad  \lambda_0 = -2J_1\,, \qquad \qquad\textrm{for}\quad    \forall c_x,  \forall c_y,  \forall c_z 
\label{L0} \\
&& J_2 > J_1\,,\quad   \lambda_0 = -2J_2\,, \qquad\qquad \textrm{for}\quad  c_x=-1,  \forall c_y,  \forall c_z
\nonumber 
\end{eqnarray}

\begin{figure}[tb]
\centering
\includegraphics[width=0.6\textwidth]{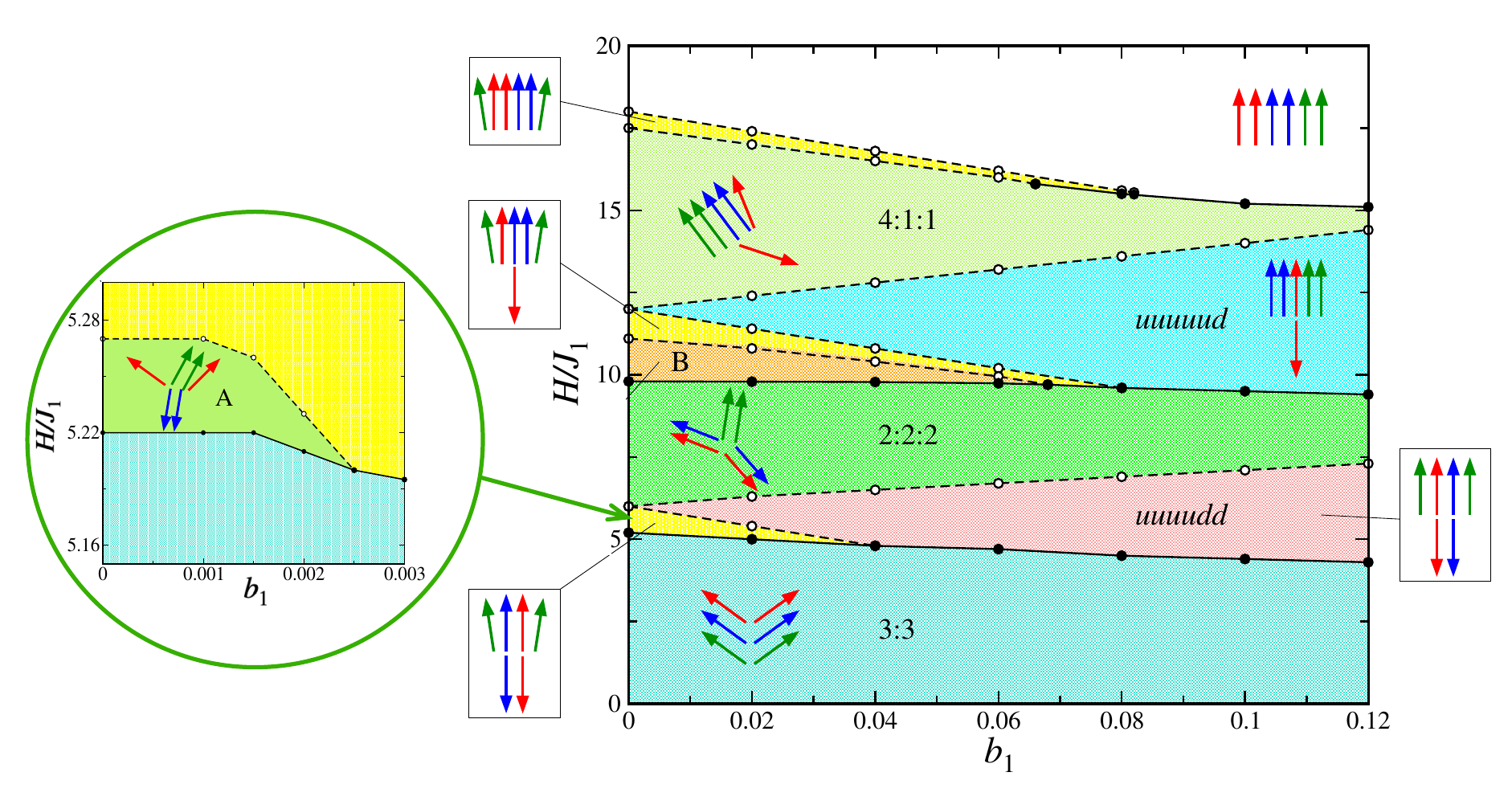} 
\caption{Detailed $H$--$b$ phase diagram of the single octahedron model for a fixed ratio 
of biquadratic exchange constants $b_2/b_1 = 1/2$.}
\label{fig:Hbdetail}
\end{figure}

For $J_2 < J_1$, the smallest eigenvalue of the exchange matrix has minima at  $(0,0,q_z)$ and 
equivalent cubic directions.  Such a degeneracy corresponds in real space to  independent 
rotations for  the  parallel square planes of spins, as discussed in the main text. 
For $J_2>J_1$,  the octahedron energy given by Eq.~(2) in the  main text is minimized for $H=0$  by  
$ \bm{S}_{d_n}=0$.
Accordingly, chains of antiparallel spins are formed on the $J_2$ bonds along three cubic directions.  
The octahedral constraint $ \bm{S}_{\rm tot}=0$ is satisfied, leaving chains completely uncorrelated. In momentum space, this corresponds to planes of degenerate lowest energy modes, as found above in Eq.~(\ref{L0}).
At the fully frustrated point $J_2 = J_1$ located between the two regimes,
the smallest eigenvalue $\lambda_0(\bm{q})$ is completely flat in
the entire Brillouin zone. 
Such a situation corresponds to completely local zero-energy modes, which reside on square voids of the corner-shared octahedral lattice.

\section{Phase diagram for a single octahedron}

The phase diagram of a single octahedron with bilinear-biquadratic interactions is discussed in
 detail in the main text and in Appendix A. Here we present further details on the appearance of the low symmetric
 phase A (2\,:\,2\,:\,1\,:\,1). This phase occupies a significant  region in the Monte Carlo phase diagram of the fully frustrated fcc model, see Fig.~2 in the main text. The phase A also appears in the $H$--$b$ diagram of the single octahedron model, albeit in a tiny region, as shown in Figure~\ref{fig:Hbdetail}.

\section{Monte Carlo data}

The Monte Carlo simulations have been performed on cubic clusters with $N = L^3$ spins and linear sizes $L = 6$--24. 
The  size of  simulated clusters is limited by the first-order nature of most phase transitions,
with pronounced hysteresis features in  field scans at low $T$.
A hybrid Monte Carlo algorithm has been employed with the canonical Metropolis sweep over the lattice followed by 10 microcanonical overrelaxation sweeps. 
For every $(T,H)$ point, up to $5\cdot10^5$ Metropolis steps have been used for equilibration, with subsequent 
1--$2\cdot 10^6$ hybrid Monte Carlo steps for measurements. In addition, averaging over 100
independent Monte Carlo runs have been performed for each temperature-field point.

\begin{figure}[thb]
\centerline{
\includegraphics[width=0.29\textwidth]{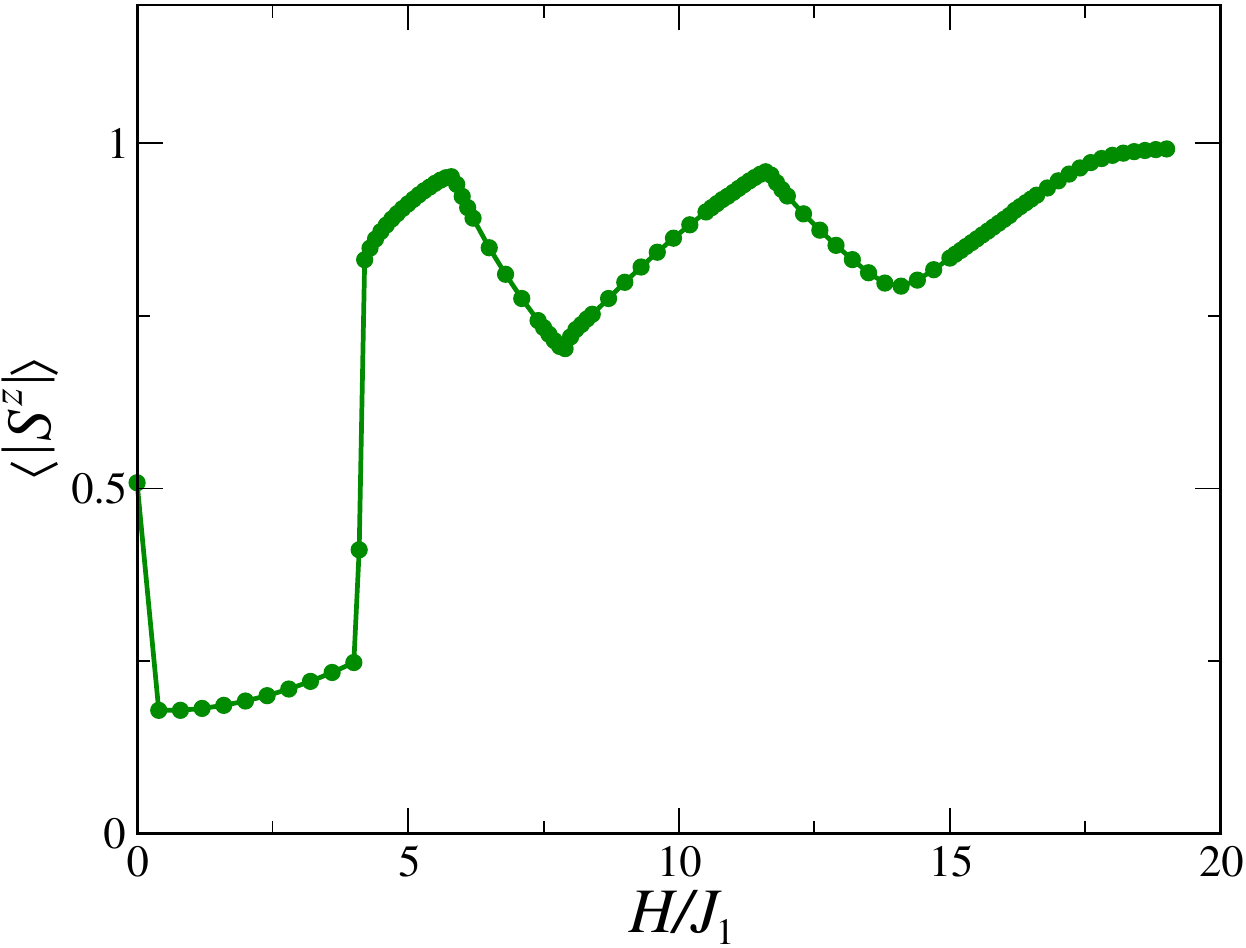} 
\hskip 3mm
\includegraphics[width=0.32\textwidth]{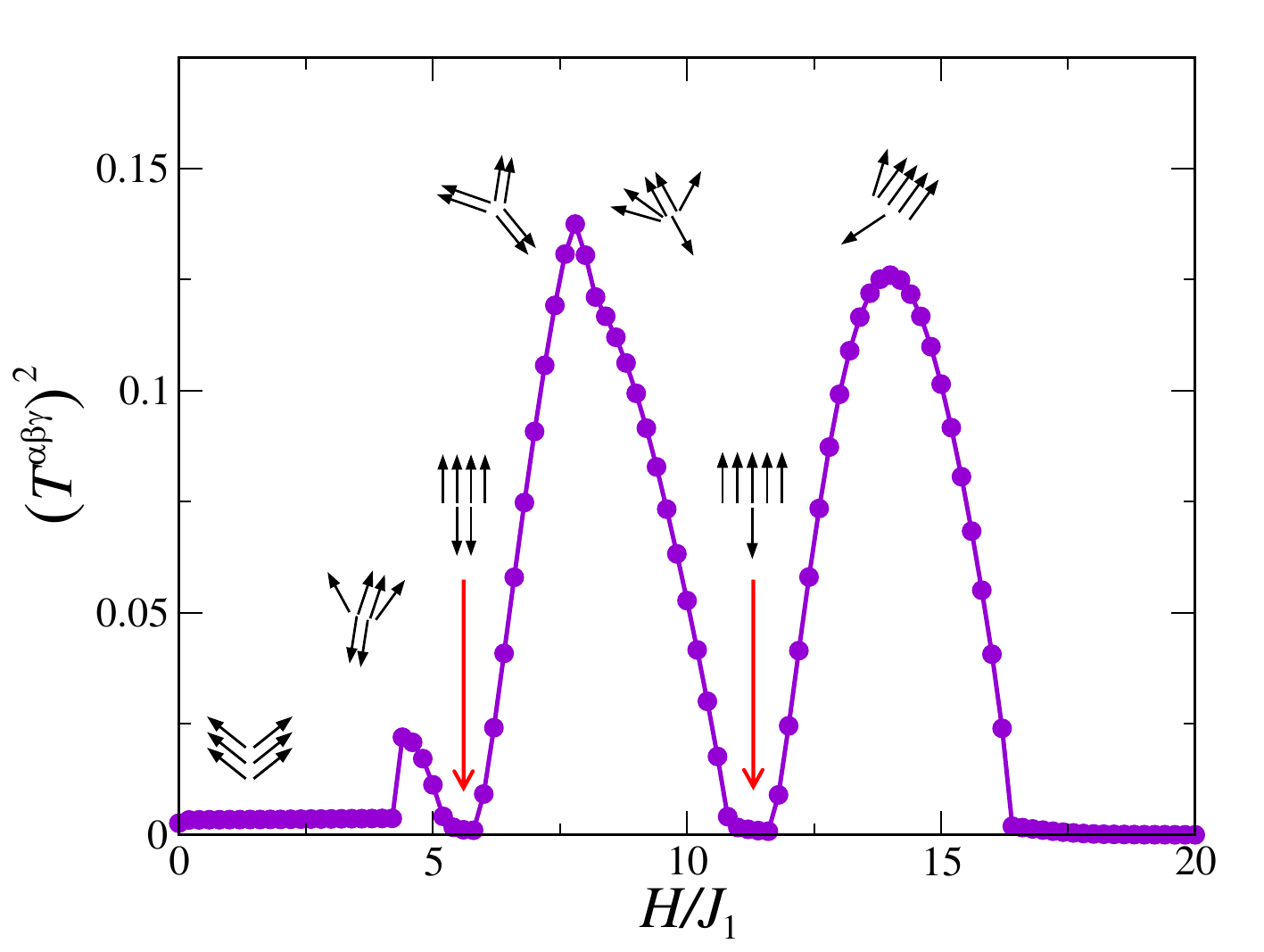} 
\hskip 3mm
\includegraphics[width=0.29\textwidth]{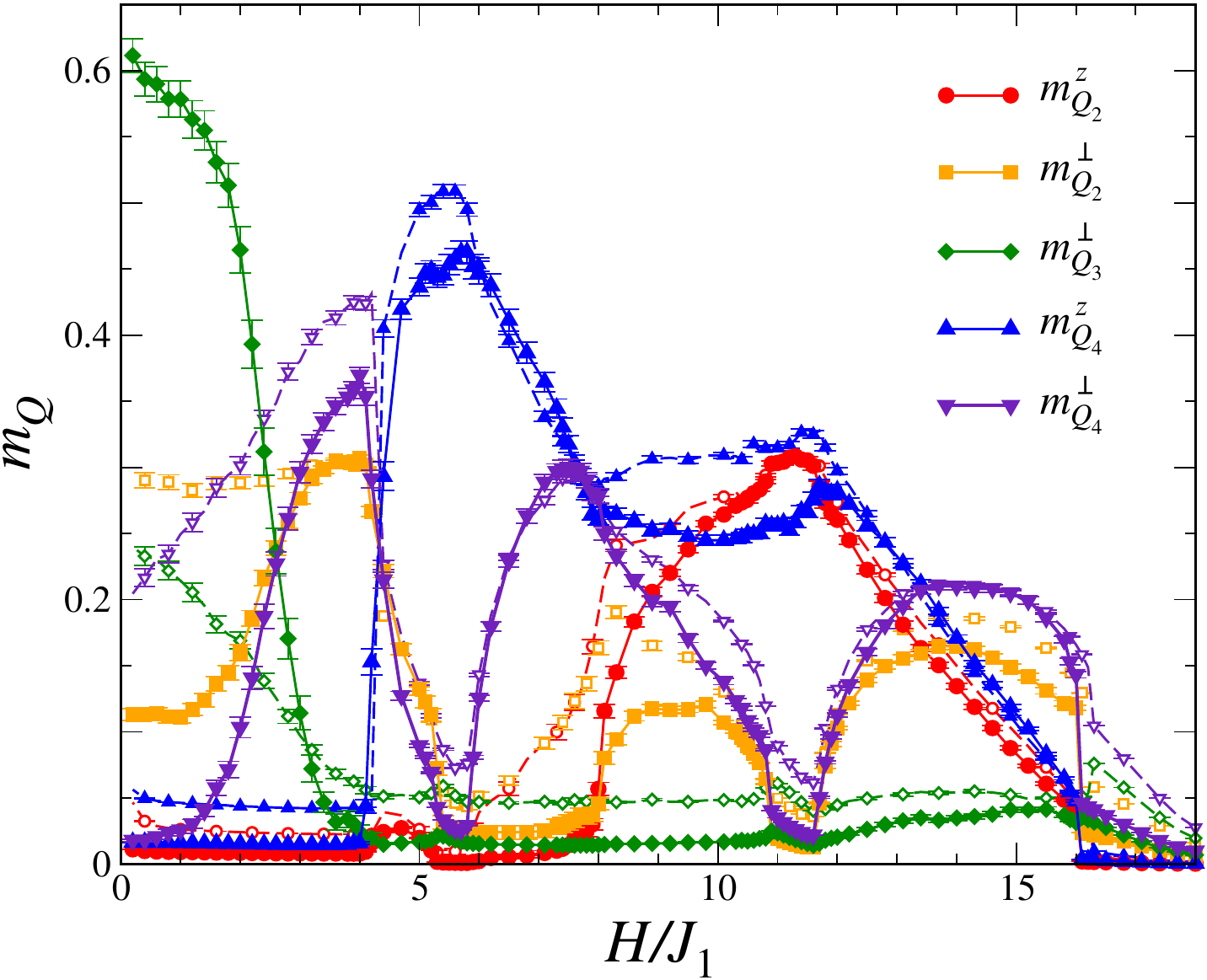} 
}
\caption{Magnetic field scans at $T/J_1=0.02$. Left panel:  the absolute longitudinal  
($z$) component of spin $\langle |S_z|\rangle$. Central panel: the transverse octupole order parameter.
Right panel: the antiferromagnetic order parameters for the three relevant wavectors. Large filled symbols 
are obtained for the spin cluster with $L=24$. Small open symbols denote the same components $m_{\bm{Q}}$
as  the corresponding large symbols, but for the  $L=12$ cluster.
}
\label{fig:SMC}
\end{figure}

Figure~\ref{fig:SMC} presents the Monte Carlo results obtained on the $L=24$ cluster  
for an average absolute value of the longitudinal spin component
\begin{equation}
\langle |S^z| \rangle
 = \frac{1}{N}\sum_i  |S_i^z|
\end{equation}
and the transverse octupole order parameter
\begin{equation}
 T^{\alpha\beta\gamma} = \frac{1}{N}\sum_i  S_i^\alpha S_i^\beta S_i^\gamma \qquad
\textrm{with}\quad  \alpha,\beta,\gamma = x,y 
\label{Tabc}
\end{equation}
versus applied magnetic field ($\bm{H}\parallel \bm{\hat{z}}$) at $T/J_1=0.02$.
Both quantities have been used to distinguish magnetic structures for the field induced
states. In particular, the fact that $\langle |S^z| \rangle\to 1$ in the two plateau regions unambiguously
indicates the development of collinear $uuuudd$ and $uuuuud$ states.
On the other hand, the square of the octupole order (\ref{Tabc}) vanishes for 
noncoplanar states with the residual $C_2$ rotational symmetry about the field direction
and acquires finite values once this symmetry is broken. In the mean-field approximation,
the  $C_2$ symmetric states exist in narrow magnetic field regions  below 
the collinear $uuuuuu$, $uuuuud$, and $uuuudd$ states, see Fig.~\ref{fig:Hbdetail}.
They are clearly absent in the Monte Carlo field scans, which demonstrate direct first-order transitions
from the collinear plateau states into various noncoplanar states.

Finally, we present in the right panel of Fig.~\ref{fig:SMC} the field dependence of the Fourier harmonics $m^\alpha_{\bm{Q}}$ of the antiferromagnetic order parameter for lattices with $L=12$ and 24. The variation of amplitudes for two clusters
allows to conclude about survival of the harmonics in the thermodynamic limt $L\to\infty$.

\end{document}